\documentclass[runningheads]{llncs}

\usepackage{subfigure}
\usepackage{graphicx}
\usepackage{amsmath}
\usepackage{bm}
\usepackage{url}
\usepackage{stmaryrd}
\usepackage{balance}
\usepackage{amssymb}
\usepackage{pgfplots}
\usepackage{pifont}
\usepackage{float}
\usepackage{epsfig}
\usepackage{booktabs}
\usepackage{pgffor}
\usepackage{tikz}
\usepackage{multirow}
\usepackage{mathrsfs}
\usepackage{bbm}
\usepackage{helvet}
\usepackage{courier}
\usepackage{threeparttable}
\usepackage{algpseudocode} 
\usepackage{arydshln}
\usepackage{mathrsfs}
\usepackage{rotating} 
\usepackage{adjustbox}
\usepackage{makecell}
\usepackage{diagbox} 
\usepackage{enumitem}
\usepackage{subfloat}
\usepackage[sort,numbers]{natbib}

\title{Reviewing Developments of Graph Convolutional Network Techniques for Recommendation Systems}

\author{
   Haojun Zhu $^1$, 
   Vikram Kapoor $^2$,
    Priya Sharma $^2$
   \\
}
\authorrunning{A. Patel et al.}
\institute{Institute of Advanced Scientific Research, Bangalore, Karnataka, India\\
	\email{aishaPetel21@iasr.org} \and
    Department of Information Science, Shivaji University\\ 
	\email{vikramkapoor20@shivaji.edu.in, prysharma@shivaji.edu.in}}

\begin{document}
\maketitle

\begin{abstract}
The Recommender system is a vital information service on today's Internet. 
Recently, graph neural networks have emerged as the leading approach for recommender systems. 
We try to review recent literature on graph neural network-based recommender systems, covering the background and development of both recommender systems and graph neural networks. 
Then caategorizing recommender systems by their settings and graph neural networks by spectral and spatial models, we explore the motivation behind incorporating graph neural networks into recommender systems. 
We akso analyze challenges and open problems in graph construction, embedding propagation and aggregation, and computation efficiency. 
This guides us to better explpre the future directions and developments in this domain.
\end{abstract}

\section{Introduction}

recommendation systems play a pivotal role in today's digital landscape, facilitating personalized content delivery and enhancing user experiences across various platforms. 
These systems analyze user preferences and behavior to predict and recommend items, such as movies, products, or articles, that are likely to be of interest. With the rapid development of recent computer science and artificial intelligence techniques, recommendation systems recieve great attention in recent years~\cite{DBLP:conf/sigir/HeDeng20,DBLP:journals/computer/KorenBV09,DBLP:conf/sigir/Wang0WFC19,koren2009matrix}.

Traditional recommendation systems have relied on methods like collaborative filtering~\cite{rendle2009bpr,koren2009mf,rendle2010factorization} that leverages the preferences and behavior of users to generate recommendations. 
Collaborative filtering relies on user interactions and similarities between users to make predictions about their preferences for items.
One of the advantages of collaborative filtering is its ability to make recommendations without requiring explicit information about the items themselves. Instead, it relies on the collective behavior and preferences of users. However, collaborative filtering does face challenges, such as the cold start problem (difficulty recommending items for new users or items with limited interactions) and the sparsity of user-item interaction data.
In recent years, collaborative filtering techniques have been combined with other approaches, including advanced machine learning techniques such as matrix factorization (MF)~\cite{koren2009mf} or factorization machine~\cite{rendle2010factorization} and deep learning such as neural networks~\cite{he2017neural,guo2017deepfm,cheng2016wide,ying2018graph}, to enhance the accuracy and effectiveness of recommender systems.

In recent years, there has been a paradigm shift with the emergence of graph neural networks (GNNs) as a state-of-the-art approach.
The integration of collaborative filtering with graph neural networks (GNNs)~\cite{hamilton2017graphsage,kipf2017semi} is an example of how cutting-edge technologies are being applied to address the complexities of recommendation scenarios, providing more personalized and accurate suggestions to users.
The motivation behind incorporating GNNs into recommendation systems lies in their ability to exploit the inherent graph structure of user-item interactions. 
GNNs leverage the relationships between users and items to provide more accurate and context-aware recommendations~\cite{wu2022graph}. 
This is particularly beneficial in scenarios where user preferences are influenced not only by the inherent qualities of items but also by the relationships and interactions within the user-item graph.

One of the key advantages of using GNNs in recommendation systems is their capacity to handle sparse and incomplete data~\cite{wu2022graph}. Traditional collaborative filtering methods may struggle when dealing with missing or incomplete user-item interactions. GNNs, on the other hand, can effectively propagate information through the graph, inferring potential connections and preferences even in situations where explicit interactions are limited.
Despite the promising advancements facilitated by GNNs, challenges persist in the areas of graph construction, embedding propagation/aggregation, model optimization, and computation efficiency. Addressing these challenges is crucial to unlocking the full potential of GNNs in recommendation systems. 
In this paper, we introduce these contents with careful discussions.

\section{Backgrounds}

\subsection{Recommendation Systems}

Recommendation systems have undergone a remarkable evolution, transforming from early collaborative filtering and content-based filtering methods to the sophisticated algorithms that shape digital experiences today. 
Collaborative filtering, whether user-based or item-based, established itself by leveraging user interactions and similarities to predict preferences. User-based collaborative filtering identifies users with similar tastes, recommending items liked by these users to the target user. 
In contrast, item-based collaborative filtering recommends items based on the similarity between items themselves. 
These approaches laid the groundwork for hybrid methods, which amalgamate collaborative and content-based filtering to harness the strengths of both, effectively mitigating challenges like the cold start problem and data sparsity.

The collaborative filtering paradigm is built on the assumption that users with similar preferences in the past will continue to share similar tastes in the future. 
However, these methods face challenges when dealing with sparse and incomplete data, a common issue in real-world scenarios. As a response to these challenges, the field has seen a paradigm shift with the integration of advanced technologies, such as matrix factorization~\cite{koren2009mf} or factorization machine~\cite{rendle2010factorization} and deep learning, to enhance the accuracy and efficiency of recommendation systems.
Particularly in industrial applications, where adherence to real-world system engineering demands is essential, recommendation systems are consistently divided into three distinct stages: \textit{matching}, \textit{ranking}, and \textit{re-ranking}. 

In the initial "Matching" stage, the objective is to efficiently select candidate items from an extensive pool, often numbering in the millions or billions, significantly reducing the overall scale. Due to the large volume of data and strict latency constraints in online serving, the utilization of complex algorithms like very deep neural networks is impractical, necessitating the use of concise models. Real-world recommender systems frequently integrate multiple matching channels, each employing distinct models such as embedding matching, geographical matching, popularity matching, and social matching \cite{covington2016deep,kang2019candidate}.

Following the "Matching" stage, in the "Ranking" stage, candidate items from various channels are consolidated into a unified list and scored by a singular ranking model. This model ranks items based on scores, enabling the incorporation of more sophisticated algorithms and considerations of rich features such as user profiles and item attributes to enhance recommendation accuracy \cite{kang2018self,song2019autoint,lian2018xdeepfm,chen2022learning}. The primary challenge in this stage lies in designing models capable of capturing intricate feature interactions.

The "Re-ranking" stage is necessary after the "Ranking" stage to address additional criteria like freshness, diversity, and fairness \cite{pei2019personalized}. This stage may involve the removal or reordering of items to align with business needs. The primary concern is considering various relationships among the top-scored items, as the proximity of similar or substitutable items can lead to information redundancy, necessitating adjustments in the displayed order \cite{ai2018learning,zhuang2018globally}.

Recommendation settings encompass a diverse array of techniques and methodologies aimed at enhancing the precision and relevance of content suggestions in various domains. Among these, three key aspects stand out: \textit{user-item collaborative filtering}, \textit{user-user social regularization}, and \textit{item-item side information supplementation}.
Specifically, in user-item collaborative filtering, the primary focus is on leveraging the collective preferences and behaviors of users to make personalized recommendations~\cite{he2017neural,DBLP:conf/sigir/Wang0WFC19,sedhain2015autorec,guo2020fastif,kang2018self}. This method establishes connections between users and items, allowing the system to infer user preferences based on historical interactions. The collaborative filtering paradigm can be further categorized into user-based and item-based approaches. User-based collaborative filtering identifies users with similar preferences, recommending items liked by those with analogous tastes. Item-based collaborative filtering, on the other hand, recommends items based on their similarity to items previously favored by the user. These collaborative approaches are foundational to many recommendation systems, providing a robust framework for generating accurate and contextually relevant suggestions.

User-user social regularization~\cite{cialdini2004social,mcpherson2001birds} introduces a social dimension to the recommendation setting by incorporating social network information. 
Beyond individual preferences, this approach considers the relationships and interactions between users within a social network. 
Users are connected based on social ties, and the recommendation algorithm takes into account the preferences and activities of users within one's social circle. 
By integrating social regularization, the system aims to improve the accuracy of recommendations by considering the influence of friends or connections on an individual's preferences. 
This adds a layer of personalization that extends beyond direct user-item interactions.

As for item-item side information supplementation, using knowledge graphs into recommendation systems broadens the recommendation landscape by incorporating additional information about items beyond user interactions~\cite{DBLP:conf/sigir/YangHuang22,wang2019kgat,wang2019knowledge,chen2021attentive,ckan,kgat}. This supplementary information can include textual descriptions, categorical tags, or any other relevant metadata associated with items. By considering these side information attributes, the recommendation system gains a more comprehensive understanding of item characteristics. This enrichment facilitates a more nuanced matching of user preferences with item features, leading to improved recommendation accuracy. The incorporation of side information becomes particularly valuable in scenarios where the content of items plays a significant role in user choices.

\subsection{Graph Convolution Networks}

A notable advancement in recent years is the incorporation of graph convolutional networks (GCN)~\cite{kipf2017semi,velivckovic2017gat,berg2018gcmc,zhang2019heterogeneous,fout2017protein,feng2019hypergraph} into recommendation systems. 
GCNs excel in modeling complex relationships within graph-structured data, making them well-suited for capturing intricate user-item interactions. Motivated by the high-order connectivity, structural properties, and enhanced supervision signals offered by GCNs~\cite{zhou2020graph,wu2020comprehensive}., researchers have explored their application in recommendation scenarios. 
The utilization of GCNs introduces a more comprehensive understanding of user preferences, considering not only the intrinsic qualities of items but also the relationships and interactions within the user-item graph.

When dealing with regular Euclidean data such as images or texts, Convolutional Neural Networks (CNNs) prove highly effective in extracting localized features. However, their application to non-Euclidean data, like graphs, necessitates a degree of generalization to address situations where the objects of operation (such as pixels in images or nodes in graphs) lack a fixed size. In the context of Graph Representation Learning (GRL), the primary objective is to generate low-dimensional vectors representing graph nodes, edges, or subgraphs, capturing the intricate connection structures within graphs. For instance, pioneering works such as DeepWalk~\cite{perozzi2014deepwalk} employ the SkipGram~\cite{mikolov2013efficient} approach on randomly generated paths through graph walks, aiming to learn node representations. This technique forms the foundation for graph-based neural networks.

The synergy of CNNs and GRL has led to the development of various Graph Convolutional Networks (GCNs), designed to distill structural information and derive high-level representations from graph data. These networks integrate the strength of CNNs in extracting localized features with the graph-aware representation learning provided by GRL. By combining these methodologies, GCNs can effectively handle non-Euclidean data structures and capture the relational intricacies present in graphs. This integration has proven particularly valuable in tasks where understanding the structural dependencies within graph data is essential, such as social network analysis, molecular chemistry, or the topic of this paper recommendation systems~\cite{zhou2020graph,wu2020comprehensive,zhang20explainable,chen2023wsfe}.

A notable example illustrating this integration is the use of DeepWalk, where the learned node representations from random walks on graphs are further refined using CNNs. This fusion allows the model to discern complex patterns and dependencies within graph structures, enhancing its ability to generalize and make accurate predictions. As the field of graph-based neural networks continues to evolve, the amalgamation of CNNs and GRL stands as a potent approach for effectively processing and understanding non-Euclidean data, offering promising avenues for advancements in various application domains.

The success of GCN-based recommenders can be attributed to three key perspectives: \textit{structural data}, \textit{high-order connectivity}, and \textit{supervision signal}.

Data derived from online platforms manifests in diverse forms, including user-item interactions, user profiles, and item attributes. 
Traditional recommender systems often struggle to harness this multi-modal data efficiently, typically focusing on specific sources and thereby overlooking valuable information. 
GCNs provide a unified approach by representing all data as nodes and edges on a graph. 
This not only enables the utilization of various data forms but also empowers GCNs to generate high-quality embeddings for users, items, and other features, crucial for optimizing recommendation performance.

The accuracy of recommendations hinges on capturing the similarity between users and items, with this similarity reflected in the learned embedding space. Specifically, the embedding for a user should align with the embeddings of items the user has interacted with, as well as those interacted with by users with similar preferences (the collaborative filtering effect). 
Traditional approaches often fall short in explicitly capturing this effect, primarily considering only first-order connectivity based on directly connected items. 
The collaborative filtering effect is naturally expressed as multi-hop neighbors on the graph, seamlessly integrated into learned representations through embedding propagation and aggregation.

Supervision signals, typically sparse in collected data, pose a challenge for recommender systems. 
GCN-based models address this by leveraging semi-supervised signals during the representation learning process. For instance, in an E-Commerce platform, the target behavior like purchases may be sparse compared to other behaviors. 
GCN-based models effectively incorporate multiple non-target behaviors, such as searches and adding to carts, by encoding semi-supervised signals over the graph, leading to significant improvements in recommendation performance \cite{jin2020multi}. 

In summary, GCN-based recommenders have redefined the landscape of recommendation systems by capitalizing on the richness of structural data, embracing high-order connectivity, and strategically leveraging supervision signals. These advancements contribute to their remarkable performance across diverse recommendation scenarios and pave the way for more effective and personalized content suggestions.

\section{GCN-based Recommendation Models}

\subsection{GCN Outlooks}

There are primarily three types of tasks on graphs: classification, prediction, and regression, occurring at three levels—node, edge, and subgraph. Despite the task diversity, a standard optimization procedure exists. Embeddings are mapped along with labels to formulate the loss function, and common optimizers are employed for model learning. Various mapping functions (\textit{e.g.}, MLP, inner product) and loss functions (\textit{e.g.}, pair-wise, point-wise) can be chosen based on specific tasks. For instance, in pair-wise loss functions like Bayesian Personalized Ranking (BPR)~\cite{rendle2009bpr}, discrimination between positive and negative samples is encouraged, and the formulation is as follows:

\begin{equation}
	\mathcal{L} = \sum_{p, n} -\textrm{ln}\sigma(s(p) - s(n)),
\end{equation} 

where $\sigma(\cdot)$ is the sigmoid function, $p$ and $n$ denote positive and negative samples, and $s(\cdot)$ measures the samples. In contrast, point-wise loss functions include mean square error (MSE) loss, cross-entropy loss, and others.

To illustrate GNN model optimization in link prediction and node classification, consider link prediction. The likelihood of an edge existing between nodes $i$ and $j$ is calculated based on the similarity with node embeddings in each propagation layer:

\begin{equation}
	s\left(i, j\right) = f(\{\mathbf{h}^l_i\}, \{\mathbf{h}^l_j\}),
\end{equation}

where $f(\cdot)$ denotes the mapping function. Training data $\mathcal{O} = \{(i, j, k)\}$ consists of observed positive and randomly-selected negative samples, $(i, j)$ and $(i, k)$, respectively. For node classification, node embeddings are transformed into a probability distribution representing its class:

\begin{equation}
	\mathbf{p}_i = f(\{\mathbf{h}_i^l\}),
\end{equation}

where $\mathbf{p}_i\in \mathbf{R}^{C\times 1}$ is the distribution, and $C$ is the number of classes. Training data $\mathcal{O} = \{(i, \mathbf{y}_i)\}$ is structured such that $\mathbf{y}_i\in \mathbf{R}^{C\times 1}$, and $i$ belonging to class $c$ is denoted by $\mathbf{y}_{ic} = 1$; otherwise, $\mathbf{y}_{ic} = 0$. For classification tasks, the point-wise loss function, like cross-entropy loss, is typically chosen:

\begin{equation}
	\mathcal{L} = -\sum_{(i, \mathbf{y}_i)\in \mathcal{O}} \mathbf{y}_i^T \log{\mathbf{p}_i}.
\end{equation}

\subsection{User-item Interaction with Collaborative Filtering}
A line of research~\cite{TrustWalker,BiRank,HOP-rec,RippleNet} exploits higher-order connectivity information between users and items to infer user preferences. For example, TrustWalker~\cite{TrustWalker} utilizes random walks to directly propagate preference scores. However, none of these approaches leverage such information in the embedding space.

Another relevant research avenue involves exploiting the user-item graph structure for recommendation. Previous efforts, like ItemRank~\cite{ItemRank}, employ label propagation to directly disseminate user preference scores across the graph, encouraging connected nodes to have similar labels. Recently, GRMF~\cite{rao2015collaborative} and HOP-Rec~\cite{HOP-rec} smooth node embeddings between one-hop and high-hop neighbors by introducing additional loss terms. These methods incorporate graph structure at the objective function level, distinguishing themselves from SVD++ which encodes neighborhood information in the predictive model formulation.

Apart from utilizing on-graph representation ability, another category of collaborative filtering (CF) methods considers historical items to profile a user, enriching user representations. Early works like FISM~\cite{FISM} construct a user representation via the average or weighted sum of ID embeddings of historical items. Later, SVD++ integrates such representations with the user's ID embedding as a final representation, achieving success in rating prediction. Nevertheless, historical items contribute differently to shaping a user's preference, necessitating adaptive learning of their weights. Recent works such as ACF~\cite{ACF}, NAIS~\cite{NAIS}, and DeepICF~\cite{DeepICF} introduce attention mechanisms to specify varying importance of historical items, achieving improved embeddings.

When revisiting historical interactions as a user-item bipartite graph, these improvements are attributed to encoding a user's ego network, i.e., her one-hop neighbors, into the embedding learning. Recently emerged graph neural networks (GNNs) have gained prominence in modeling graph structure, especially high-hop neighbors, to guide embedding learning~\cite{GCN,GraphSAGE}. GNNs were initially proposed for node classification on attributed graphs, where each node is described by rich input features (e.g., attributes, contents). The basic idea of GNN is to encode graph structure, as well as input features, into a better representation of each node. Early studies defined graph convolution in the spectral domain, such as Laplacian eigen-decomposition~\cite{DBLP:journals/corr/BrunaZSL13} and Chebyshev polynomials~\cite{FirstGCN}, which are computationally expensive. Later on, GraphSage~\cite{GraphSAGE} and GCN~\cite{GCN} redefined graph convolution in the spatial domain, aggregating the embeddings of neighbors to refine the target node's embedding. Due to its interpretability and efficiency, this formulation quickly became prevalent in GNNs and is widely used~\cite{DeepInf,Feng2019TOIS,zhao2019cross}.

Motivated by the strength of graph convolution, recent efforts like NGCF~\cite{NGCF}, GC-MC~\cite{GC-MC}, and PinSage~\cite{PinSage} adapted GCN to the user-item interaction graph, capturing CF signals in high-hop neighbors for recommendation.

It is worth mentioning that several recent efforts provide deep insights into GNNs~\cite{DeepInsights,ICLR19-APPNP,SGCN}. 
In particular, Wu et al.~\cite{SGCN} argue for the unnecessary complexity of GCN, developing a simplified GCN (SGCN) model by removing nonlinearities and collapsing multiple weight matrices into one. 
Another work conducted concurrently~\cite{LR-GCCF} also finds that nonlinearity is unnecessary in NGCF and develops a linear GCN model for CF. 
Recently, a light-weighted GCN architecture namely LightGCN attracts great attention as it removes unnecessary modules, such as removing all redundant parameters and retaining only the ID embeddings, making the model as simple as matrix factorization (MF).

\subsection{User-user Soical Regularization}

One of the earliest instances of a social recommender system dates back to 1997. 
In recent years, an abundance of social media platforms, such as Facebook and Twitter, has emerged, providing individuals with convenient means to communicate and express themselves. The widespread adoption of social media has resulted in an unprecedented volume of social information.
For instance, Facebook, the largest social networking site, has fostered a staggering 35 billion online friendships. 
Similarly, the most popular user on Twitter, the leading microblogging site, boasts an impressive 37,974,138 followers. The exponential growth of social media has significantly expedited the advancement of social recommender systems.

In recent years, there are lots of works exploiting user's social relations for improving the recommender system~\cite{wu2018social_collaborative, tang2013exploiting, tang2016recommendations}. 
Most of them assume that users' preference is similar to or influenced by their friends, which can be suggested by social theories such as social homophily~\cite{mcpherson2001birds} and social influence~\cite{marsden1993network}. 
According to the assumptions above, social regularization has been proposed to restrain the user embedding learning process in the latent factor based models~\cite{ma2011recommender, jamali2010matrix, ma2008sorec}. 
And TrustMF~\cite{yang2016social} model is proposed to model the mutual influence between users by mapping users into two low-dimensional space: truster space and trustee space and factorize the social trust matrix. 
By treating the social neighbors' opinion as the auxiliary implicit feedbacks of the targeted user, TrustSVD~\cite{guo2015trustsvd} is proposed to incorporate the social influence from social neighbors on top of SVD++~\cite{koren2008factorization}. 
Generally, this technique extends traditional matrix factorization methods by incorporating social regularization terms into the objective function. Social regularization encourages the learned user and item embeddings to respect the social relationships, making the recommendation model aware of the influence of social connections.

Moreover, some recent studies like~\cite{wang2017item, fan2018deep, chen2019social} and~\cite{fan2019deep, chen2019efficient, krishnan2019modular} leverage deep neural network and transfer learning or adversarial learning approach respectively, to learn a more complex representation or model the shared knowledge between social domain and item domain. 
GCNs have gained popularity in social regularized recommendation. They capture the complex relationships in a social network by learning node embeddings through iterative information aggregation from neighboring nodes. The learned embeddings can then be used for improved recommendation accuracy.

Social regularized recommendation enhances the personalization of recommendations by considering the influence of social connections. It acknowledges that users with similar social ties may share common preferences, leading to more accurate and personalized recommendations.
By incorporating social connections, the recommendation system can introduce diversity in recommendations. It can avoid the "filter bubble" problem where users are only exposed to a narrow set of items, thus promoting serendipitous discovery.
In summary, social regularized recommendation techniques offer promising avenues to enhance recommendation systems by incorporating social network information. While they bring advantages in terms of personalization and diversity, addressing challenges related to data sparsity, scalability, and privacy is crucial for their widespread and ethical adoption.

\subsection{Item-item Side Information}

Exploring Knowledge Graphs (KGs) as a form of supplementary information has garnered interest in various applications, particularly in recommender systems. Existing recommender models that incorporate KGs fall into three primary categories: path-based~\cite{Hete-cf, HINRec, MCRec, RuleRec}, embedding-based~\cite{ckbe, DKN, IKSR}, and hybrid methods~\cite{wang2018ripplenet, kgat, ckan, chen2022modeling, wang2021learning}.

Path-based methods delve into various connecting patterns among items in KGs, such as meta-paths or meta-graphs, to offer additional guidance for recommendations. The generation of these patterns relies either on path generation algorithms~\cite{HERec} or manual creation~\cite{MCRec}. While path-based methods naturally introduce interpretability and explanation into recommendations, designing such patterns can be challenging, particularly for large-scale and complex KGs. The exhaustive retrieval and generation of paths become impractical, and the selection of paths significantly impacts the final recommendation performance.

Embedding-based methods employ knowledge graph embedding (KGE) algorithms~\cite{KGE_survey} to directly utilize semantic information in KGs and enhance the representations of users and items. For instance, DKN~\cite{DKN} utilizes TransD~\cite{TransD} to jointly process KGs and learn item embeddings. However, the limitations of embedding-based methods lie in emphasizing rigorous semantic relations and neglecting user-item interactions in recommender systems. Moreover, most embedding-based methods lack support for end-to-end training.

Hybrid methods integrate path-based and embedding-based techniques, aiming to achieve state-of-the-art performance. These methods typically employ iterative information propagation under a graph neural network framework to generate entity representations for information enrichment. For example, CKAN~\cite{ckan} utilizes a heterogeneous propagation strategy along multi-hop links to encode knowledge associations for users and items.

Two models, MetaHIN and MetaKG~\cite{metahin, metakg}, leverage the power of the meta-learning paradigm, treating preference learning for each user as a single meta-learning task. However, a significant challenge arises in the computational cost associated with optimizing individual meta-learners for each user from the interaction records. Another recent model, KGPL~\cite{KGPL}, adopts graph semi-supervised learning to employ pseudo-labeling via random walk, simulating a graph to increase interaction density. Nonetheless, starting with structure exploration may be influenced by the initial state of interaction sparsity, potentially perturbing and destabilizing the model training for recommendation. These models are included in experiments for performance comparison.

\section{Conclusions}

Early works on Graph Neural Networks (GCNs) typically adopted the full-batch gradient descent algorithm, where the entire adjacency matrix was multiplied on the node embeddings during each inference step. However, this approach proved impractical for real-world recommender systems due to the potential million-level scale of nodes and edges. Notably, a recent work by Zhao et al.~\cite{zhao2022joint} even introduced a GCN-based solution for both searching and recommendation engines, elevating the demand for enhanced computational efficiency.
Addressing the limitations of full-batch methods, Hamilton et al. proposed GraphSAGE~\cite{hamilton2017graphsage}, which employs neighbor sampling and selectively updates the related sub-graph instead of the entire graph during each inference step. This sampling strategy has been adopted in several other works~\cite{chiang2019cluster,chen2018fastgcn}, reducing the computational complexity of GCNs and enhancing scalability. Ying et al.~\cite{ying2018graph} successfully applied GraphSAGE to web-scale recommender systems, efficiently computing embeddings for billions of items.
Some works~\cite{feng2022reinforcement} have introduced specific query-recall acceleration strategies to optimize the efficiency of recommendation processes. Additionally, open-source tools like PyG~\cite{fey2019fast}, DGL~\cite{wang2019deep}, and AliGraph~\cite{zhu2019aligraph} have been released, facilitating research and development in GCN-based recommendation systems.
Another way to alleviate efficiency issue is to use quantization~\cite{gray1998quantization,chen2023bipartite}, which is also widely used for fast retrieval of images~\cite{qin2020forward,lin2017towards,hashnet}, documents~\cite{li2014two}, categorical information~\cite{kang2021learning}, e-commerce products~\cite{zhang2017discrete,chen2022learning}.
A notable advancement is the introduction of PlatoGL~\cite{lin2022platogl}, a graph learning system that balances both effectiveness and scalability. PlatoGL provides a practical solution for combining large-scale GCNs with real-time recommendation—a crucial challenge in today's industrial recommendation systems.

Existing recommendation models based on Graph Neural Networks (GCNs) predominantly operate on static graphs, overlooking the inherent dynamics present in recommender systems. This limitation becomes evident in scenarios such as sequential or session-based recommendations, where user data is inherently collected in a dynamic manner. Moreover, effectively modeling dynamic user preferences emerges as a paramount challenge in these recommendation contexts.
Furthermore, the dynamic nature of recommender systems is heightened by the potential inclusion of new users, products, features, and other evolving elements within the platform. This constant evolution poses a significant challenge to conventional static Graph Neural Networks.
In response to these challenges, the research focus has shifted towards dynamic graph neural networks~\cite{li2020dynamic,ma2020streaming}. These models incorporate embedding propagation operations on dynamically constructed graphs, allowing for a more responsive and adaptable approach to the evolving nature of recommender systems.
Given the time-evolving property inherent in recommender systems, the advent of dynamic graph neural networks presents a promising research direction. These models hold considerable potential for real-world applications, offering the flexibility to capture and adapt to the dynamic complexities of user interactions, new arrivals, and changing system features. This evolution toward dynamic graph neural network-based recommendation models reflects a crucial step in enhancing the practicality and efficacy of recommender systems in dynamic, real-world scenarios.

\bibliographystyle{unsrtnat}
\bibliography{ref}

\end{document}